\documentclass{iaus}
\usepackage{graphicx}
\usepackage{natbib}

\title[Powerful explosions at $Z=0$?] 
{Powerful explosions at \boldmath $Z=0$?}

\author[Ekstr\"om \& al.]   
{Sylvia Ekstr\"om$^1$, Georges Meynet$^1$, Raphael Hirschi$^2$ \& Andr\'e Maeder$^1$}

\affiliation{$^1$Geneva Observatory, University of Geneva, Maillettes 51 - CH 1290 Sauverny, Switzerland \\[\affilskip]
$^2$University of Keele, Keele, ST5 5BG, UK}

\pubyear{2008}
\volume{255}  
\jname{Low-Metallicity Star Formation: From the First Stars to Dwarf Galaxies}
\editors{L.K. Hunt, S. Madden \& R. Schneider, eds.}

\begin{document}
\maketitle

\begin{abstract}
Metal-free stars are assumed to evolve at constant mass because of the very low stellar winds. This leads to large CO-core mass at the end of the evolution, so primordial stars with an initial mass between 25 and 85 $M_{\odot}$ are expected to end as direct black holes, the explosion energy being too weak to remove the full envelope.

We show that when rotation enters into play, some mass is lost because the stars are prone to reach the critical velocity during the main sequence evolution. Contrarily to what happens in the case of very low- but non zero-metallicity stars, the enrichment of the envelope by rotational mixing is very small and the total mass lost remains modest. The compactness of the primordial stars lead to a very inefficient transport of the angular momentum inside the star, so the profile of $\Omega(r)$ is close to $\Omega\,r^2=$ const. As the core contracts, the rotation rate increases, and the star ends its life with a fast spinning core. Such a configuration has been shown to modify substantially the dynamics of the explosion. Where one expected a weak explosion or none at all, rotation might boost the explosion energy and drive a robust supernova. This will have important consequences in the way primordial stars enriched the early Universe.
\end{abstract}

\firstsection 
\section{The first stars}
Population III (Pop III) stars occupy a key position in the history of the Universe, as a link between the pure H-He Universe at the beginning of times, and the metal-rich one we observe nowadays around us. Their evolution is important, because during their life they form the first heavy elements, but their death is also important, because this is the moment at which the newly synthesised elements are released in the surrounding medium, starting the chemical enrichment of the Universe.

While the first studies on the star formation at $Z=0$ concluded that only very massive objects could form \citep{bcl02,abn02}, later findings show that the radiative feedback of the very first stars could lead to a second generation of metal-free stars with lower initial mass \citep{omyosh03,oshabwn05,greifbr06}. Could these stars contribute to the chemical enrichment of the Universe? The consensual picture is that most of them are directly swallowed by a black hole.

\section{Evolution at \boldmath $Z=0$}

Since they are deprived of metals, the massive Pop III stars cannot rely on the CNO cycle to sustain their gravity. The $pp$-chains are not very sensitive in temperature and cannot completely halt the initial collapse. The stars continue their contraction until the central temperature is high enough to allow some carbon to be produced through the 3$\alpha$ reaction.

\begin{figure}[t]
\begin{center}
 \includegraphics[width=4.5in]{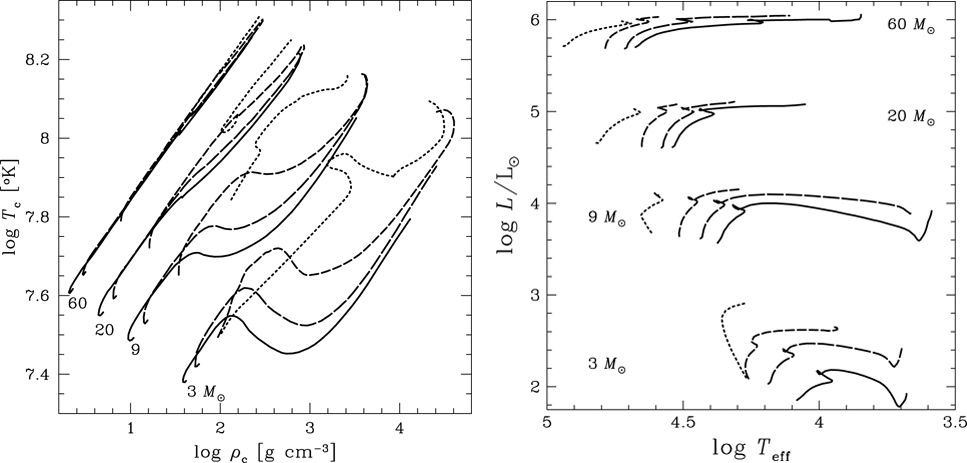} 
 \caption{Evolution during the main sequence of massive stars at various metallicities: $Z=0.020$ (solid line); $Z=0.002$ (long-dashed line); $Z=10^{-5}$ (short-dashed line); $Z=0$ (dotted line). \textsl{Left:} $\log T_{\rm c} - \log \rho_{\rm c}$ diagram. \textsl{Right:} HR diagram.}
   \label{fdhrtcrc}
\end{center}
\end{figure}

As a consequence, the main sequence (MS) occurs at much higher central temperature ($T_{\rm c}$) and density ($\rho_{\rm c}$), as shown in Fig.~\ref{fdhrtcrc} (\textsl{left}). The great compactness implies that the zero-age main sequence (ZAMS) is shifted toward higher effective temperature ($T_{\rm eff}$), as shown in Fig.~\ref{fdhrtcrc} (\textsl{right}). Without any metal, the envelope is transparent, so the stars remain in the blue side of the HR diagram throughout the whole MS.

The lack of metals has another consequence: the radiative winds are supposed to scale with the metallicity as $\dot{M}\propto(Z/Z_\odot)^\alpha$ with $\alpha=0.5$ \citep{kupu00,nuglam00,Kudr02} or 0.7-0.8 \citep{vink01,vdk05,mok07}. At $Z=0$, we thus expect no radiative winds, and most Pop III models are computed with the hypothesis of constant mass. In fact, \citet{Kudr02} shows that very low-metallicity stars close to the Eddington limit $L_{\rm Edd}=4\pi cGM/\kappa$ (with $\kappa$ the electron-scattering opacity and obvious meaning for the other quantities) are subject to a weak but non-null mass loss. \citet{vdk05} show that there is a flattening in the metallicity dependence of the WR winds below $Z\approx10^{-3}$. Also \citet{smow06} show that continuum-driven eruptive mass ejections can occur at any metallicity, even $Z=0$, leading to strong mass loss episodes like those observed in luminous blue variables (LBVs).

Even if the mass loss is non null, it is weak anyway, so the Pop III stars end their life with a large helium core ($M_\alpha$), as shown by \citet{mar01}. In such a case, the expected fate of the stars can deviate from the usual ``SN explosion with neutron star or black hole remnant'' seen at higher metallicity. \citet{hfw03} determine the fate of massive stars at various metallicities as a function of their $M_\alpha$ at the pre-supernova (preSN) stage:
\begin{itemize}
\item a Type II SN: $M_\alpha < 9\ M_\odot$
\item a BH by fallback: $9\ M_\odot \leq M_\alpha < 15\ M_\odot $
\item a direct BH: $15\ M_\odot  \leq M_\alpha < 40\ M_\odot $
\item a pulsational pair-instability followed by a SN with BH formation: $40\ M_\odot  \leq M_\alpha < 64\ M_\odot $
\item a pair-instability SN (PISN): $64\ M_\odot  \leq M_\alpha \leq 133\ M_\odot $
\item or a direct BH collapse: $M_\alpha > 133\ M_\odot $
\end{itemize}
Relating $M_\alpha$ with the initial mass of the star in the case of $Z=0$, this means that stars with a mass between 25 and 140 $M_\odot$ or above 260 $M_\odot$ will not contribute at all to the enrichment of the early Universe.

Could this picture be revised?

\begin{figure}[t]
\begin{center}
 \includegraphics[width=2.3in]{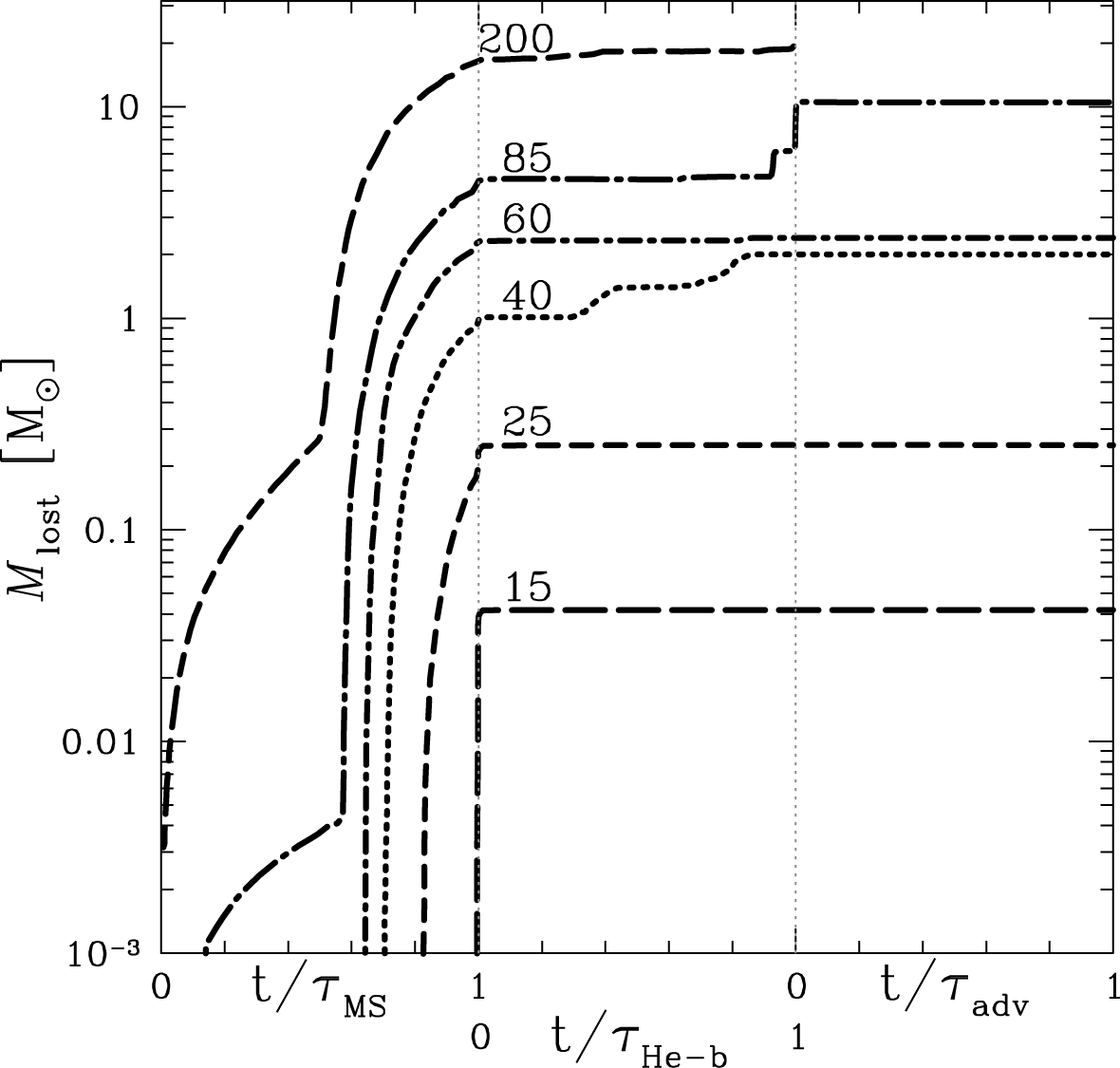} 
 \caption{Mass lost during the evolution of Pop III models of various masses \citep{eks08}: 15 $M_\odot$ (long-dash line); 25 $M_\odot$ (short-dashed line); 40 $M_\odot$ (dotted line); 60 $M_\odot$ (dot-short dashed line); 85 $M_\odot$ (dot-long dashed line); 200 $M_\odot$ (short dash-long dashed line). The $x$-axis is a temporal axis, divided in three parts: the fraction of the MS evolution, the fraction of the core He burning phase, and the fraction of the rest of the advanced phases.}
   \label{fmlost}
\end{center}
\end{figure}

\section{Rotation effects}

\citet{mem06} show that at low (but non-zero) metallicity, rotation drastically changes the evolution of massive stars. Two processes are involved:
\begin{enumerate}
\item During the MS, because of the low radiative winds, the star loses very little mass and thus very little angular momentum. As the evolution proceeds, the stellar core contracts and spins up. If a coupling exists between the core and the envelope (\textit{i.e.} meridional currents or magnetic fields), the surface may be accelerated up to the critical velocity and the star may experience a mechanical mass loss due to the centrifugal acceleration. The matter is launched into a decretion disc \citep{Ow05}, which may be dissipated later by the radiation field of the star.
\item Rotation induces an internal mixing that depends on the gradient of the rotational rate $\Omega$. After H exhaustion, the core contracts and the envelope expands, stretching the $\Omega$-gradient. The mixing becomes strong and enriches the surface in heavy elements. Rotation also favours a redward evolution after the MS, allowing the star to spend more time in the cooler part of the HR diagram, where mass loss is increased. The outer convective zone dives deep inside the star and dredges up freshly synthesised heavy elements. The surface metallicity is dramatically enhanced (by a factor of 10$^6$ for a 60 $M_\odot$ at $Z_{\rm ini}=10^{-8}$). The opacity of the envelope is increased, and the radiative winds may thus be drastically enhanced.
\end{enumerate}
\citet{mem06} show that at $Z=10^{-8}$, while a non-rotating  60 $M_\odot$ model loses only 0.27 $M_\odot$, a similar model computed with an initial velocity $\upsilon_{\rm ini}= 800$ km s$^{-1}$ loses 36 $M_\odot$, \textsl{i.e.} 60 \% of its initial mass.

Now what happens if the metallicity is $Z=0$ strictly?
\begin{figure}[t]
\begin{center}
 \includegraphics[width=4.5in]{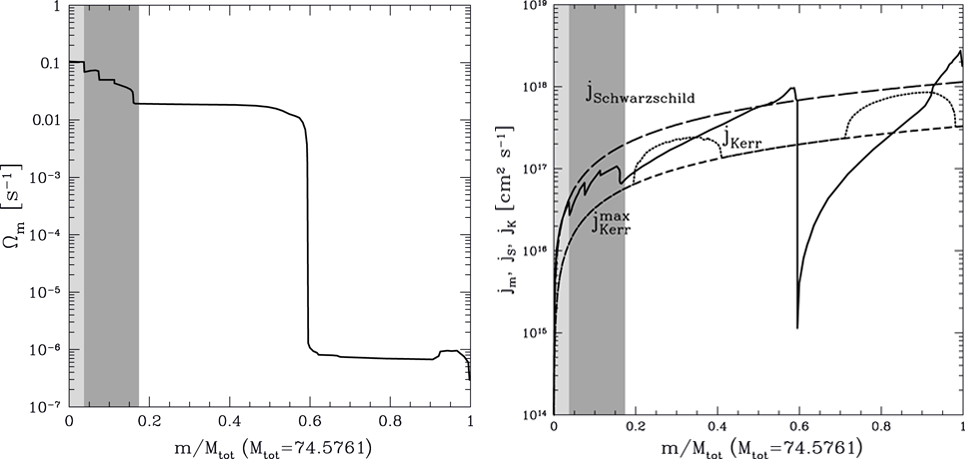} 
 \caption{Pre-supernova structure of a Pop III 85 $M_\odot$ model The light grey shaded area marks the border of the remnant, and the dark grey that of the CO core. \textsl{Left:} $\Omega$ profile. \textsl{Right:} specific angular momentum profile. $j_\mathrm{Schwarzschild}=\sqrt{12}Gm/c$ (long-dashed line) and $j_\mathrm{Kerr}^\mathrm{max}=Gm/c$ (short-dashed line) is the minimum specific angular momentum necessary for a non-rotating and a maximally-rotating black hole, respectively. $j_{\rm Kerr}$ (dotted line) is the minimum specific angular momentum necessary to form an accretion disc around a rotating black hole.}
   \label{fj85}
\end{center}
\end{figure}

\citet{eks08} show that the first process (the mechanical mass loss during the MS) is slightly reduced because the meridional circulation is weak, so the star almost evolves in a regime of local angular momentum conservation $\Omega\,r^2=$ const, and reaches the critical velocity late in its MS evolution. The second process (the enrichment of the surface) barely occurs at $Z=0$. This can be explained by the fact that some He is burnt already during the MS: at H exhaustion, the star is hot enough to move on to core He burning without any structural readjustment, and the He burning occurs mainly in the blue side of the HR diagram. The stretching of the $\Omega$-gradient does not occur, and the outer convective zone, if any, remains thin. Most of the post-MS evolution occurs with a surface metallicity much lower than 10$^{-6}$. 

As shown in Fig.~\ref{fmlost}, most of the mass is lost during the MS, and it amounts at most to 10\% of the initial mass.

\section{Explosion}
A direct consequence of the evolution close to the local angular momentum conservation is that the models keep a high angular momentum content in the core throughout their whole evolution. Figure~\ref{fj85} (\textsl{left}) shows that the angular velocity of the iron core of a 85 $M_\odot$ model at the pre-supernova stage is more than 5 orders of magnitude higher than the surface velocity. On the right panel of the same figure, we see that in the innermost parts, the specific angular momentum of the same model is higher than that needed to create an accretion disc around a rotating BH.

According to \citet{nom03}, there are indications that the most energetic supernovae, the hypernovae (HN), could be due to stellar rotation. Rotation is supposed to boost the explosion energy of a collapsing star by various mechanisms:
\begin{itemize}
\item it drives an anisotropy of the neutrinos sphere \citep{shim01,kot03},
\item it lowers the critical neutrinos luminosity (the luminosity at which the stalled shock revives) \citep{yama05},
\item it modifies the gravity \citep{burr05},
\item it changes the geometry of the mass accretion, making it aspherical \citep{burr05},
\item it generates vortices, which dredge up heat and increase the neutrinos luminosity \citep{burr05}.
\end{itemize}
The models presented here are massive at the time of their death, they have heavy cores and they almost kept their entire envelope, but they have fast rotating iron cores, and therefore their final collapse could present a highly aspherical geometry. It would be extremely interesting to perform numerical simulations of the collapse of these rotating models so we could check whether the energy released in the collapse is sufficient to overcome the high gravitation of the star and drive a successful explosion. For the time being, we can only say that we expect rotating Pop III stars to undergo a stronger explosion than what is commonly admitted for non-rotating Pop III stars of this mass range.

\section{Conclusions}
The lack of metals affects strongly the stellar evolution at $Z=0$. One of the main result is to reduce dramatically the radiative mass loss. This means that rotating stars will keep all their angular momentum until the end of their evolution. At the pre-supernova stage, they will present a fast-rotating iron core, so we expect the collapse to be highly aspherical. This may boost the explosion energy and allow a strong supernova explosion where only a weak one or none was expected.

Note that the inclusion of magnetic fields could alter these results by providing the core-envelope coupling which is lacking in Pop III stars. However, in that case, a strong mass loss would be expected \citep{kauai08}, so the evolution of the star would anyway be strongly affected with regard to the consensual picture.

\bibliographystyle{cup}
\bibliography{ekstrom_iaus255}
\end{document}